\documentclass{aastex62}

\usepackage{multirow}
\graphicspath{{./}{figures/}}

\submitjournal{ApJ}

\shorttitle{Hydrogen Emission in Type II White-light Solar Flares}
\shortauthors{Proch\'{a}zka et al.}

\begin{document}

\title{Hydrogen Emission in Type II White-light Solar Flares}

\correspondingauthor{Ond\v{r}ej Proch\'{a}zka}
\email{oprochazka01@qub.ac.uk}

\author[0000-0003-4215-5062]{Ond\v{r}ej Proch\'{a}zka}
\affiliation{Astrophysics Research Centre\\
Queen's University Belfast \\
Northern Ireland, UK}

\author[0000-0002-7695-4834]{Aaron Reid}
\affiliation{Astrophysics Research Centre\\
Queen's University Belfast \\
Northern Ireland, UK}

\author[0000-0002-7725-6296]{Mihalis Mathioudakis}
\affiliation{Astrophysics Research Centre\\
Queen's University Belfast \\
Northern Ireland, UK}






\begin{abstract}
Type II WLFs have weak Balmer line  emission and no Balmer jump. 
We carried out a set of radiative hydrodynamic simulations to understand how the hydrogen radiative losses vary  with the electron beam parameters and more  specifically with the low energy cutoff. Our results have revealed that for low energy beams, the excess flare Lyman emission diminishes with increasing low energy cutoff as the energy deposited into the top chromosphere is low compared to the  energy deposited into the deeper layers. Some Balmer excess emission is  always present and is driven primarily by direct heating from the beam with a minor contribution from Lyman continuum backwarming. 
The absence of Lyman excess emission in electron beam models with high low energy cutoff is a prominent spectral signature of type II WLFs.
\end{abstract}



\section{Introduction} \label{sec:intro}
White-light flares (WLF) have attracted considerable attention since the Carrington event in 1859 when brightenings lasting for several minutes were observed in the solar photosphere (\citealt{Carrington:1859aa}). Early spectroscopic observations have shown that hydrogen emission dominates the excess spectrum of WLF with the Balmer lines in particular increasing considerably above the quiescent level  (\citealt{Svestka:1966ac}, \citealt{Machado:1974aa}). Early observations used the area of the emitting chromosphere as seen in H$\alpha$ together with the intensity of the emission to classify flares. The H$\alpha$ line is still routinely used to image the upper layer of the chromosphere that is greatly affected during the flare event.  The first multi-channel flare observations covering the Lyman $\alpha$ (Ly$\alpha$) that showed excess emission during a solar flare were obtained with the L.P.S.P. instrument (stands for Laboratoire de Physique Stellarie et Plan\'{e}taire, \citealt{Bonnet:1978aa}) on board the OSO-8 satellite (\citealt{Maran:1975aa,Lemaire:1984aa}). The continuous monitoring of solar irradiance in Ly$\alpha$ commenced in 2006 with the EUV Sensor (EUVS, \citealt{Viereck:2007aa}) on board the Geostationary Environmental Operational Satellite (GOES). The Extreme ultraviolet Variability Experiment (EVE; \citealt{Woods:2012aa}) on-board the Solar Dynamics Observatory (SDO)  allowed the regular acquisition of flare spectra covering the Lyman lines and continuum. The analysis of solar flares showed a rapid increase in Ly$\alpha$ during the impulsive phase (\citealt{Milligan:2016aa}) consistent with the H$\alpha$ line behaviour. \cite{Machado:2018aa} used the EVE to study the departure coefficients and the colour temperature of six X-class flares. They found a rapid increase in the Lyman continuum during the impulsive phase originating in a thin layer where the electron density exceeded 10$^{13}$ cm$^{-3}$.

The existence of two spectrally distinct types of WLFs was first reported by \cite{Machado:1986aa}. Their conclusion was based on the analysis of flaring spectra in the near-UV and visible range  (\citealt{Hiei:1982aa,Neidig:1983aa,Boyer:1985aa}) where they identified rare events with extraordinarily weak and narrow Balmer lines and without a Balmer jump. They named such events type II WLFs but were not able to explain theoretically the difference between these events and flares with strong and broad Balmer lines (type I WLF). \cite{Ding:1999aa} proposed a scenario different from the standard flare model that incorporated an energy release in the lower layers of the solar atmosphere to explain the suppressed hydrogen emission in the type II WLFs. Their interpretation included an initial decline in the WL continuum just before the flare onset and was also  elaborated by \cite{Litvinenko:1999aa} and \cite{Chen:2001aa}. \cite{Allred:2005ab} compared the atmospheric response due to electron beam heating of $10^{10}$ and $10^{11}$ erg cm$^{-2}$ s$^{-1}$ and found that the initial decline in continuum can be caused by non-thermal hydrogen ionization. They found that for the lower beam flux the decline lasts longer and therefore is more likely to be detected during the observations. \cite{Matthews:2003aa} carried out a large study of 59 WLF observed with the Yohkoh spacecraft. However, the lack of spectral information did not allow them to identify with certainty the atmospheric height where the observed emission originated and were unable to distinguish between type I and type II WLFs. 
Despite these effort, the exact definition of type II WLFs remains unclear. While it is largely accepted that these events are characterised  by suppressed Balmer line emission, some of their other features are questionable due to the lack of sufficient observational diagnostics. For example, the time lag between the WL emission, hard X-ray (HXR) emission and microwave emission is thought to be a feature of the type II WLFs (\citealt{Fang:1995aa}). However, \cite{Prochazka:2017aa,Prochazka:2018aa} presented a multi-instrument study of a type II WLF and found no temporal mismatch between WL, HXR and $\gamma$-ray emissions. Their work showed that type II WLFs are consistent with the standard flare model, and their spectral signatures can be explained with low energy particle beams with a high value for the low energy cutoff. Such beams are able to leave the upper chromosphere relatively undisturbed and deposit their energy into the deeper layers of the atmosphere. 

In this paper we use the radiative hydrodynamic code RADYN to analyze the hydrogen emission in flares that are driven by electron beams with parameters that are representative of type I and type II WLFs. We use the models of \cite{Prochazka:2018aa} that identified the best set of electron beam parameters that can recreate the observational signatures of the X1-class flare on 14 June 2014. The selected electron beam-driven models are then used to evaluate the radiative losses in both Lyman and Balmer transitions. 

\section{Flare modelling}
The RADYN code (\citealt{Carlsson:1992aa,Carlsson:1995aa,Carlsson:1997aa,Allred:2015aa}) is a one-dimensional radiative hydrodynamic code that can be used to study the interaction of particle beams with the solar atmosphere. It uses the Fokker-Planck formalism (\citealt{McTiernan:1990aa}), which takes into account the beam energy losses due to Coulomb collisions and pitch-angle diffusion when incorporating relativistic effects. RADYN includes a six level hydrogen atom, a nine level helium atom and a six level calcium atom. A return current has also been included in the simulations.
We generated a set of models that simulate the conditions in the solar atmosphere during weak to intense WLFs. The beam fluxes used had values of $3\times 10^9$, $1\times 10^{10}$ and $3\times 10^{10}$ erg cm$^{-2}$ s$^{-1}$ while the low energy cutoff E$_C$ covered the parameter space where flare values are usually found ($20 - 120$ keV, \citealt{Warmuth:2016ab}). The spectral index $\delta$ was equal to 3 for all models. The beams were applied continuously and the outputs were analysed at $t=20$ s. This value is consistent with the X-ray analysis of the best observed type II WLF to date (\citealt{Prochazka:2018aa}). 
The initial atmosphere used in this work has the transition region placed at a height of 1200 km above the photospheric floor and has a coronal temperature of 3 MK at 10 Mm (QS.SL.HT loop described in \citealt{Allred:2015aa}). The beams were injected at the top of a half-loop with a Gaussian distribution  with a half width at half max of 23.5$^\circ$.
Line synthesis was carried out using the RH code (\citealt{Uitenbroek:2001aa}) incorporating partial redistribution which is particularly important for resonance line profiles. We used the 20-level hydrogen atom and the 6-level calcium atom to produce synthetic spectra for the Lyman and Balmer line and continuum diagnostics. A VOIGT profile was used for Ca II K line, Lyman $\gamma$ the higher order Lyman lines. The Ca II H, Lyman $\alpha$ and $\beta$ profiles were modelled in PRD. The Balmer lines had profiles of type VOIGT\_VCS\_STARK, which incorporates the unified Stark effect theory (\citealt{Kowalski:2017aa}).
The spectra were synthesized by setting a minimal spectral resolution of 0.05 nm.

\section{Results}

\begin{figure}
    \centering
    \includegraphics[width=0.49\linewidth]{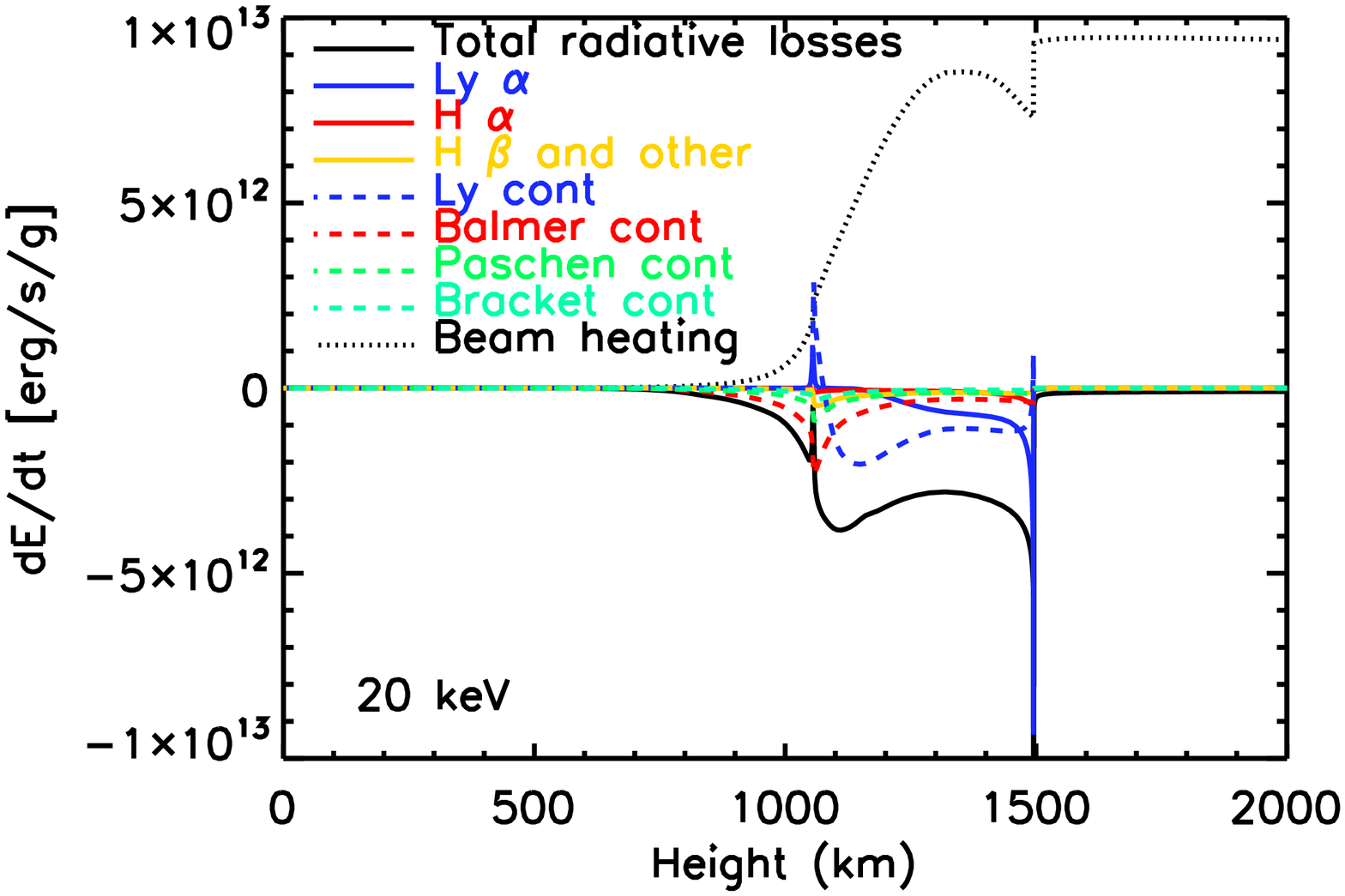}
    \includegraphics[width=0.49\linewidth]{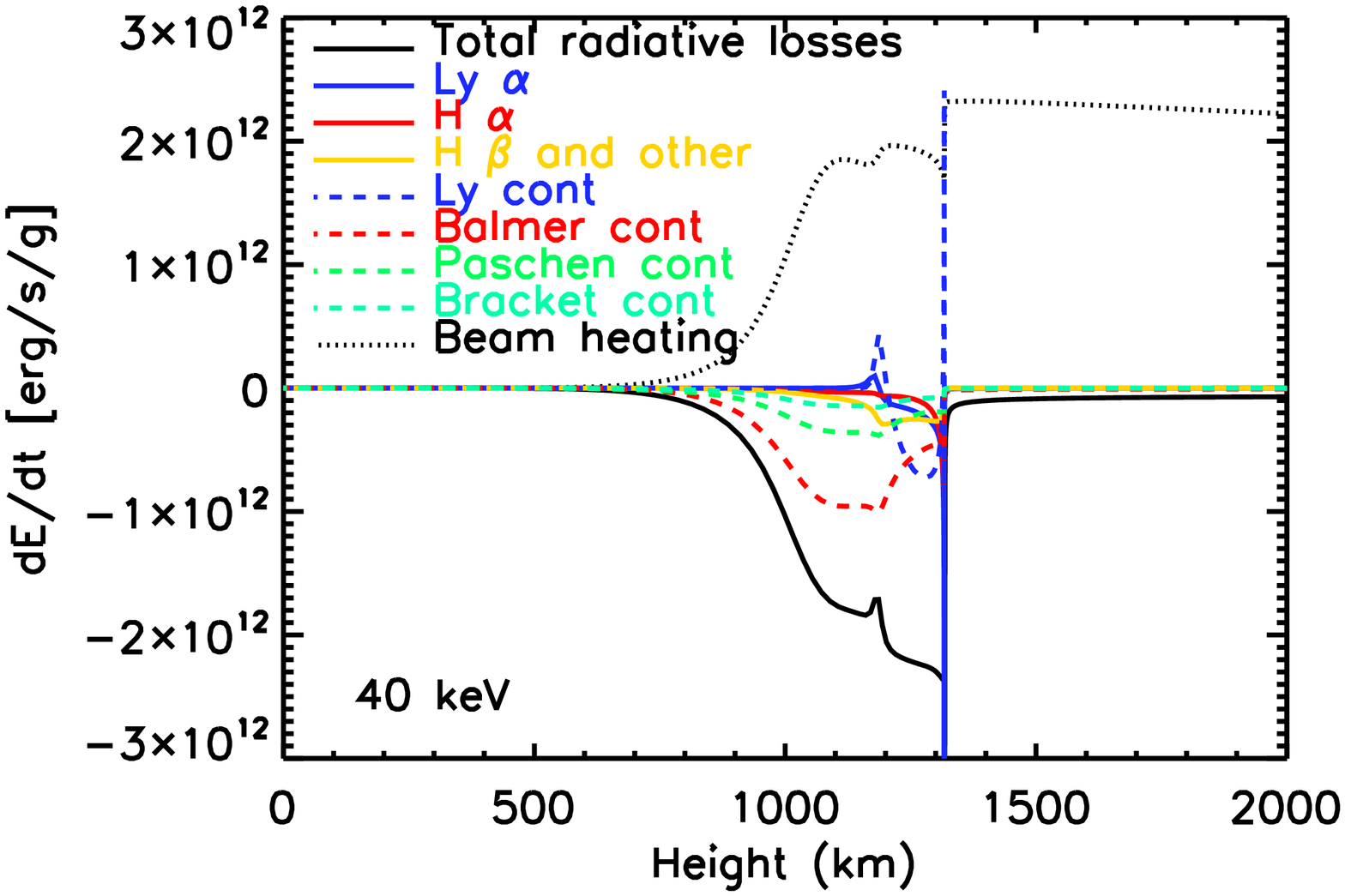}\\
    \includegraphics[width=0.49\linewidth]{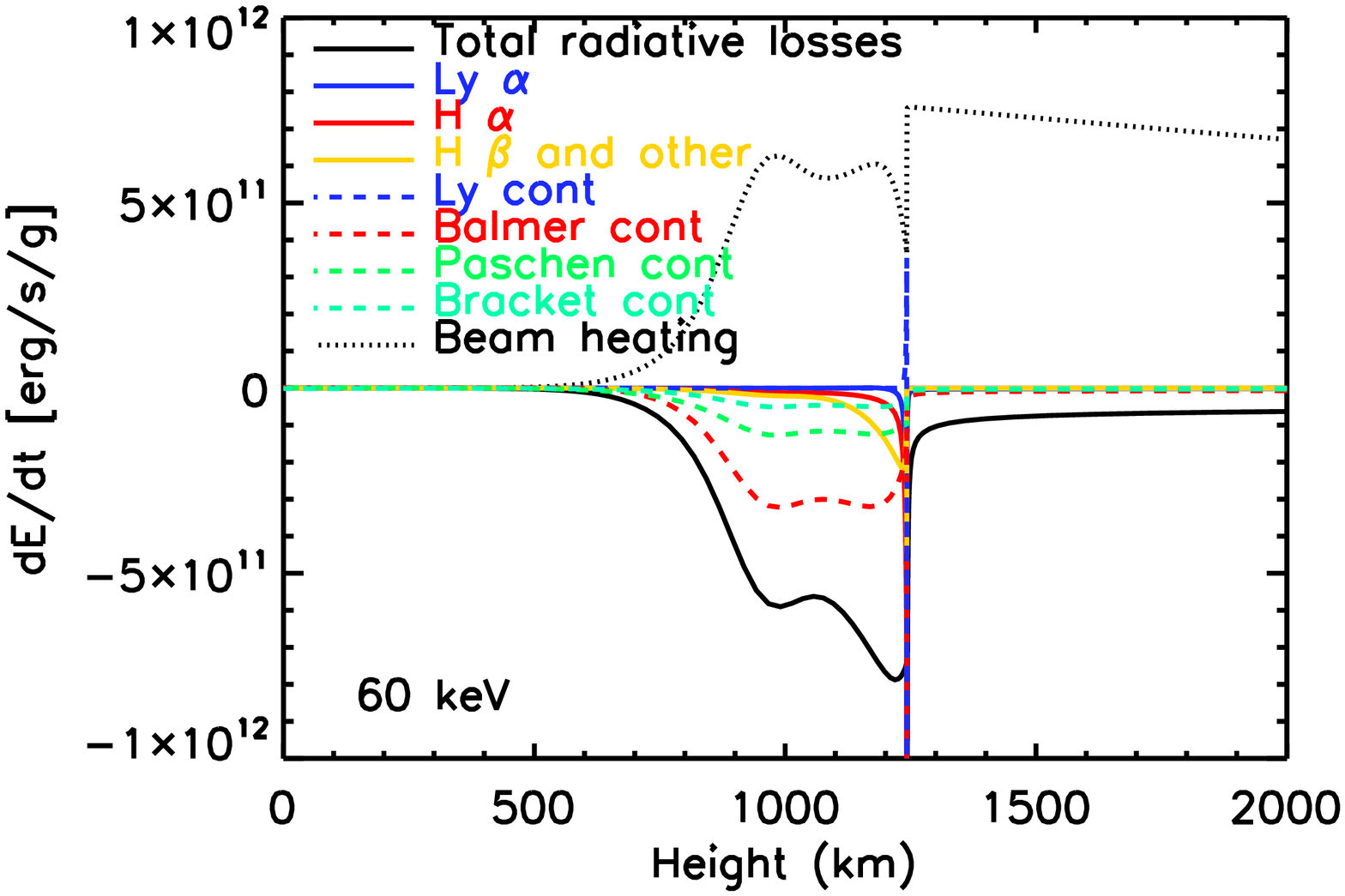}
    \includegraphics[width=0.49\linewidth]{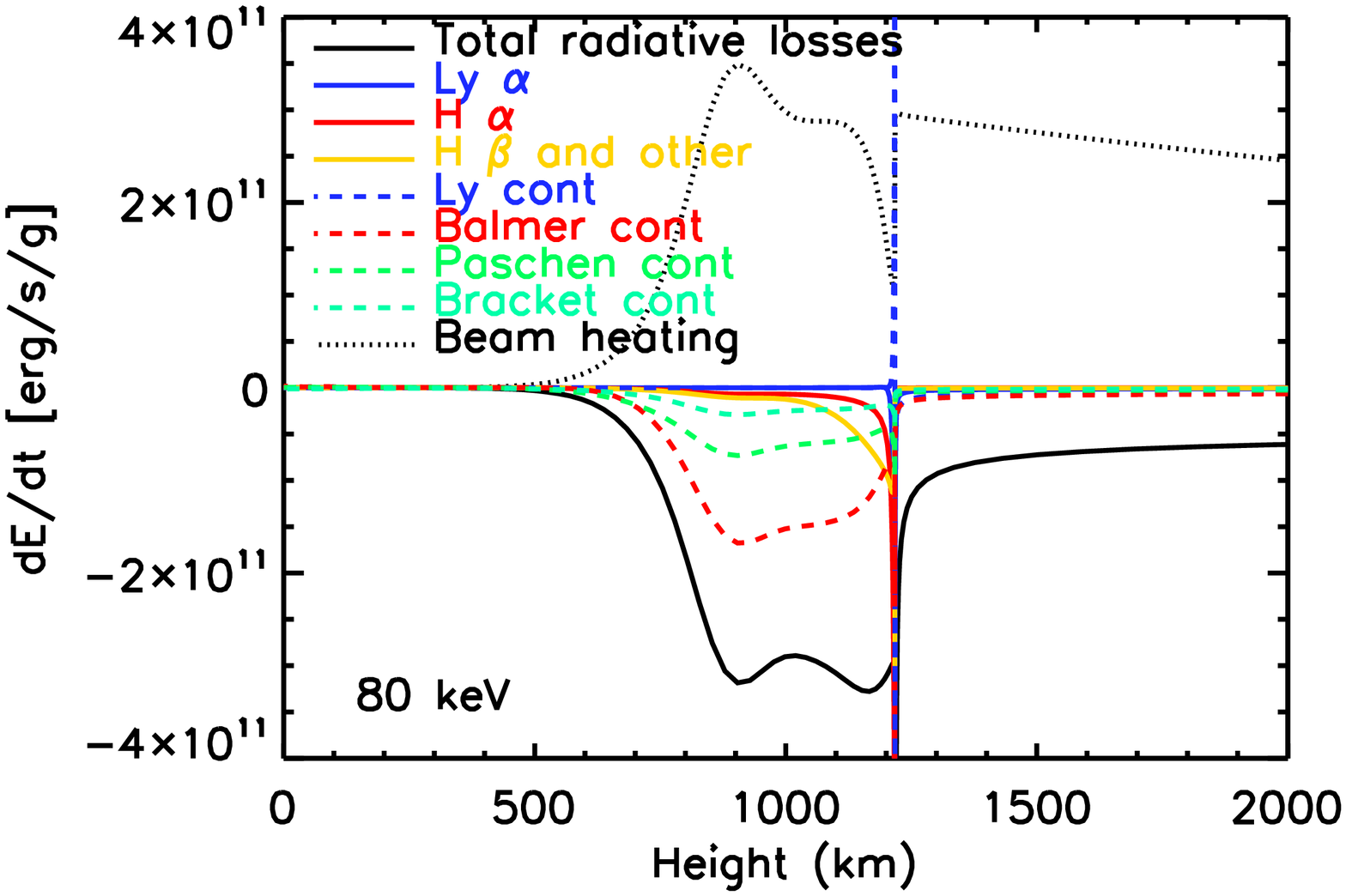}\\
    \includegraphics[width=0.49\linewidth]{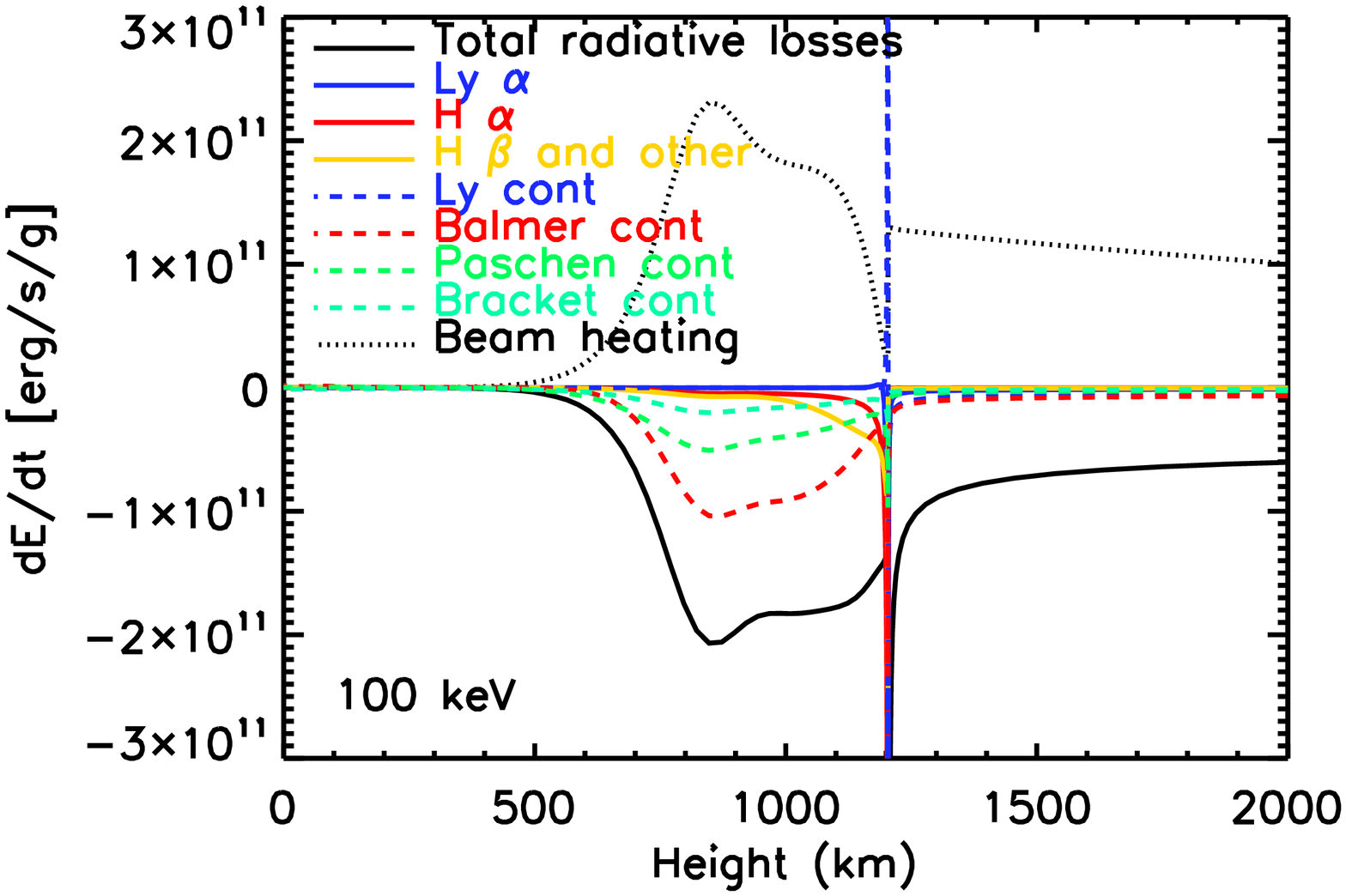}
    \includegraphics[width=0.49\linewidth]{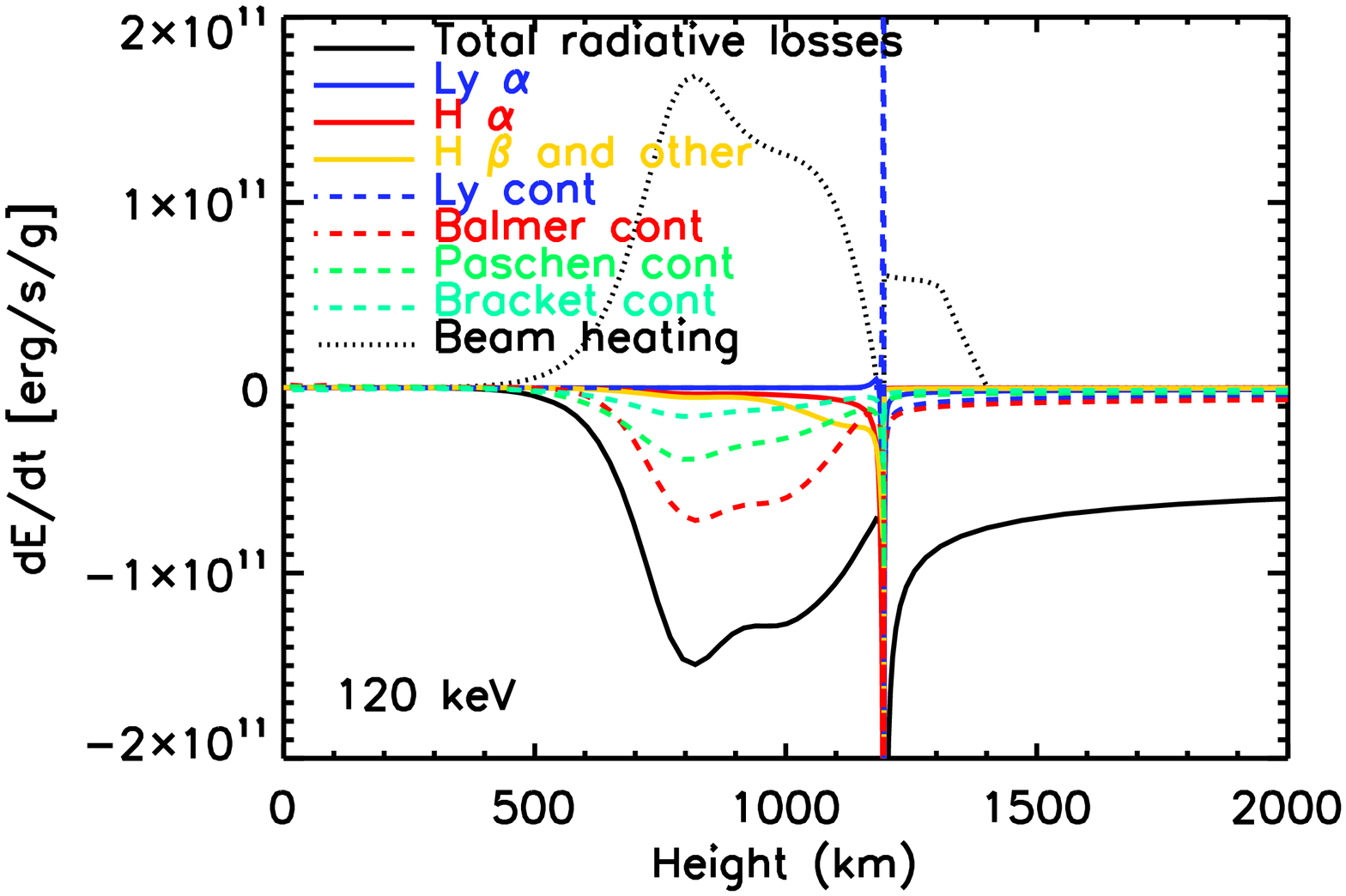}
    \caption{Radiative losses during the flare electron beam precipitation for E$_C$ in the range of 20 - 120 keV. The beam flux is equal to $3\times10^{9}$ erg cm$^{-2}$ s$^{-1}$. A positive (negative) $\frac{dE}{dt}$ corresponds to absorption (emission). The profiles are taken 20 seconds into the simulations.}
    \label{fig:1}
\end{figure}

Figure \ref{fig:1} shows the radiative losses and energy input per unit mass across the atmosphere due to the most important hydrogen transitions in models with the beam flux equal to $3\times10^{9}$ erg cm$^{-2}$ s$^{-1}$. The figure shows that beams with a low E$_C$ (20 keV) disrupt the top chromosphere, shift the transition region to heights of up to 1500 km and produce emission in the Lyman continuum and Lyman $\alpha$. The figures also show a peak of absorption in the Lyman continuum at the same height where we can see the maximum of radiative losses in the Balmer and the higher order continua. This means that the Lyman continuum irradiated downwards contributes to hydrogen ionization in the deeper layers. The radiative losses for E$_C$ greater than 40 keV do not show any such Lyman emission, but they still show the emission in the higher order continua that are clearly driven by the particle beam precipitation.

Table \ref{tab:1} shows the gross radiative losses in the atmosphere due to the free-bound Lyman and Balmer transitions (LyC and BaC, respectively) and the bound-bound Ly$\alpha$ and H$\alpha$ transitions respectively. The radiative losses were integrated only over the chromosphere and the temperature minimum region - between the height of 300 km above the photospheric floor and the height where the temperature begins to exceed $10^5$ K. These numbers do not include any possible reabsorption outside these  atmospheric heights, therefore do not quantify the radiative output of the atmosphere as seen by an observer. The table shows that there is an overall stronger emission for particle beams with higher flux with the LyC showing the greatest variation. In models with E$_C=20$ keV the LyC losses reach values of $10^{19} - 10^{20}$ erg cm s$^{-1}$ g$^{-1}$ however, models with high E$_C$ show that the LyC remains at the quiet level ($\sim 10^{16}$ erg cm s$^{-1}$ g$^{-1}$). The situation is the same for the Ly$\alpha$, where the quiet level reaches $\sim10^{17}$ erg cm s$^{-1}$ g$^{-1}$. 
The radiative losses listed in Table \ref{tab:1} show that when the beam flux is increased, the maximum radiative losses in the studied lines and continua occur at higher values of E$_C$. With increasing E$_C$ it is the LyC that reaches the peak of the radiative losses first and then the Balmer continuum. This is due to the greater penetration depth of the beams with higher E$_C$ and the fact that the LyC originates higher in the atmosphere than the BaC. 
\begin{deluxetable}{cc|c|c|c|c|c|c}[h]
    \tablecaption{Atmospheric heaating and radiative losses due to the hydrogen transitions in electron beam-driven models. The radiative losses and the beam heating, as shown in Figure \ref{fig:1}, were integrated over the temperature minimum region and chromosphere. The first row shows the radiative losses for the quiet atmosphere.}  \label{tab:1}
\tablehead{ &&   E$_C$ & Beam heating & Losses in LyC & Losses in BaC & Losses in Ly$\alpha$ & Losses in H$\alpha$ \\
 && (keV) & (erg cm s$^{-1}$ g$^{-1}$) & (erg cm s$^{-1}$ g$^{-1}$) & (erg cm s$^{-1}$ g$^{-1}$) & (erg cm s$^{-1}$ g$^{-1}$) & (erg cm s$^{-1}$ g$^{-1}$) }
    \startdata
 && & 0 &  2.13e+16 &  1.27e+16 &  2.98e+17 &  1.41e+17 \\ 
\hline    
 \multirow{6}{*}{\rotatebox{90}{F = 3$\times$10$^{9}$}} & \multirow{6}{*}{\rotatebox{90}{erg cm$^{-2}$ s$^{-1}$}} &     20 &  3.15e+20 &  5.63e+19 &  3.39e+19 &  2.29e+19 &  5.09e+18 \\
&& 40 &  6.52e+19 &  5.99e+18 &  2.94e+19 &  3.08e+18 &  2.61e+18 \\
&& 60 &  2.39e+19 &  1.47e+16 &  1.21e+19 &  4.29e+17 &  1.09e+18 \\
&& 80 &  1.35e+19 &  1.79e+16 &  6.40e+18 &  3.33e+17 &  7.03e+17 \\
&& 100 &  8.97e+18 &  1.84e+16 &  3.97e+18 &  3.08e+17 &  5.27e+17 \\
&& 120 &  6.54e+18 &  1.99e+16 &  2.68e+18 &  3.03e+17 &  4.27e+17 \\
\hline
\multirow{6}{*}{\rotatebox{90}{F = 1$\times$10$^{10}$}} & \multirow{6}{*}{\rotatebox{90}{erg cm$^{-2}$ s$^{-1}$}} &    20 &  1.03e+21 &  1.04e+20 &  6.67e+19 &  4.26e+19 &  5.97e+18 \\
&& 40 &  4.42e+20 &  8.58e+19 &  7.20e+19 &  2.49e+19 &  5.73e+18 \\
&& 60 &  1.65e+20 &  4.01e+19 &  5.77e+19 &  1.34e+19 &  4.71e+18 \\
&& 80 &  7.98e+19 &  1.93e+19 &  4.01e+19 &  6.66e+18 &  3.54e+18 \\
&& 100 &  4.52e+19 &  3.67e+18 &  2.91e+19 &  2.58e+18 &  2.45e+18 \\
&& 120 &  2.92e+19 &  4.98e+16 &  1.94e+19 &  7.12e+17 &  1.59e+18 \\
\hline
\multirow{6}{*}{\rotatebox{90}{F = 3$\times$10$^{10}$}} & \multirow{6}{*}{\rotatebox{90}{erg cm$^{-2}$ s$^{-1}$}} &   20 &  1.35e+21 &  1.35e+20 &  9.20e+19 &  2.89e+19 &  4.47e+18 \\
& &  40 &  1.59e+21 &  1.58e+20 &  1.34e+20 &  4.59e+19 &  6.98e+18 \\
& &  60 &  8.02e+20 &  1.21e+20 &  1.38e+20 &  4.00e+19 &  7.09e+18\\
& &  80 &  4.19e+20 &  8.36e+19 &  1.15e+20 &  3.52e+19 &  6.81e+18 \\
& & 100 &  2.42e+20 &  6.80e+19 &  8.89e+19 &  2.78e+19 &  6.14e+18 \\
& & 120 &  1.52e+20 &  5.35e+19 &  7.14e+19 &  2.08e+19 &  5.47e+18 \\
  \enddata
\end{deluxetable}

\begin{figure}[h]
    \centering
    \includegraphics[angle = 90, width=0.6\linewidth]{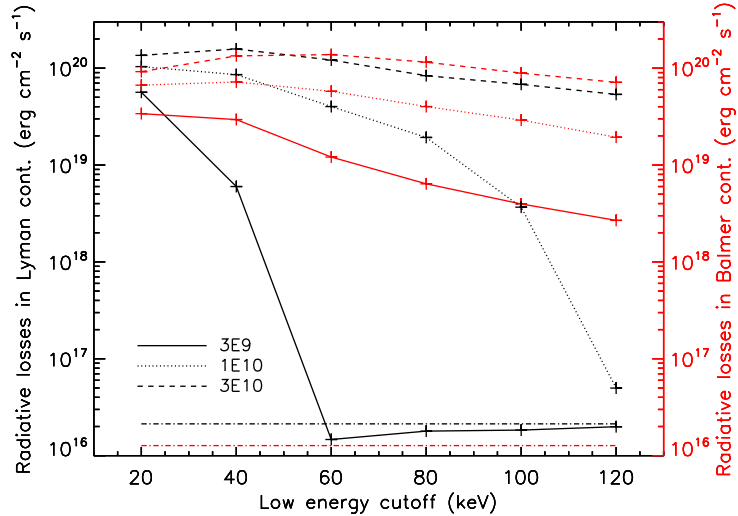}
    \caption{The gross radiative losses between the photosphere and the transition region due to the Lyman and Balmer free-bound transitions as a function of E$_C$. The legend distinguishes between the beam flux. The dot-and-dashed lines mark the level of the radiative losses with no beam applied.}
    \label{fig:4}
\end{figure}

\begin{figure}
    \centering
    \includegraphics[width=0.49\linewidth]{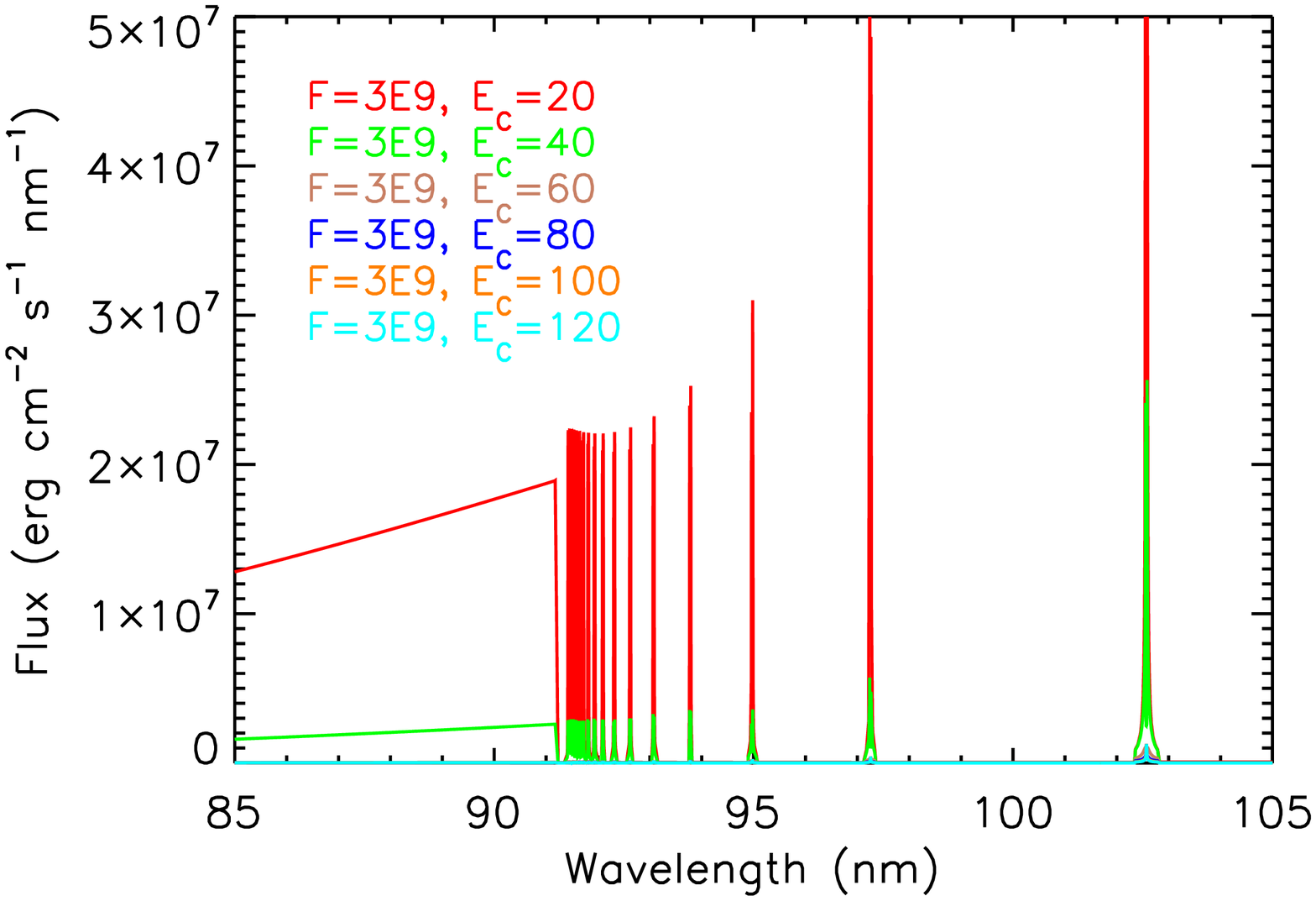} \includegraphics[width=0.49\linewidth]{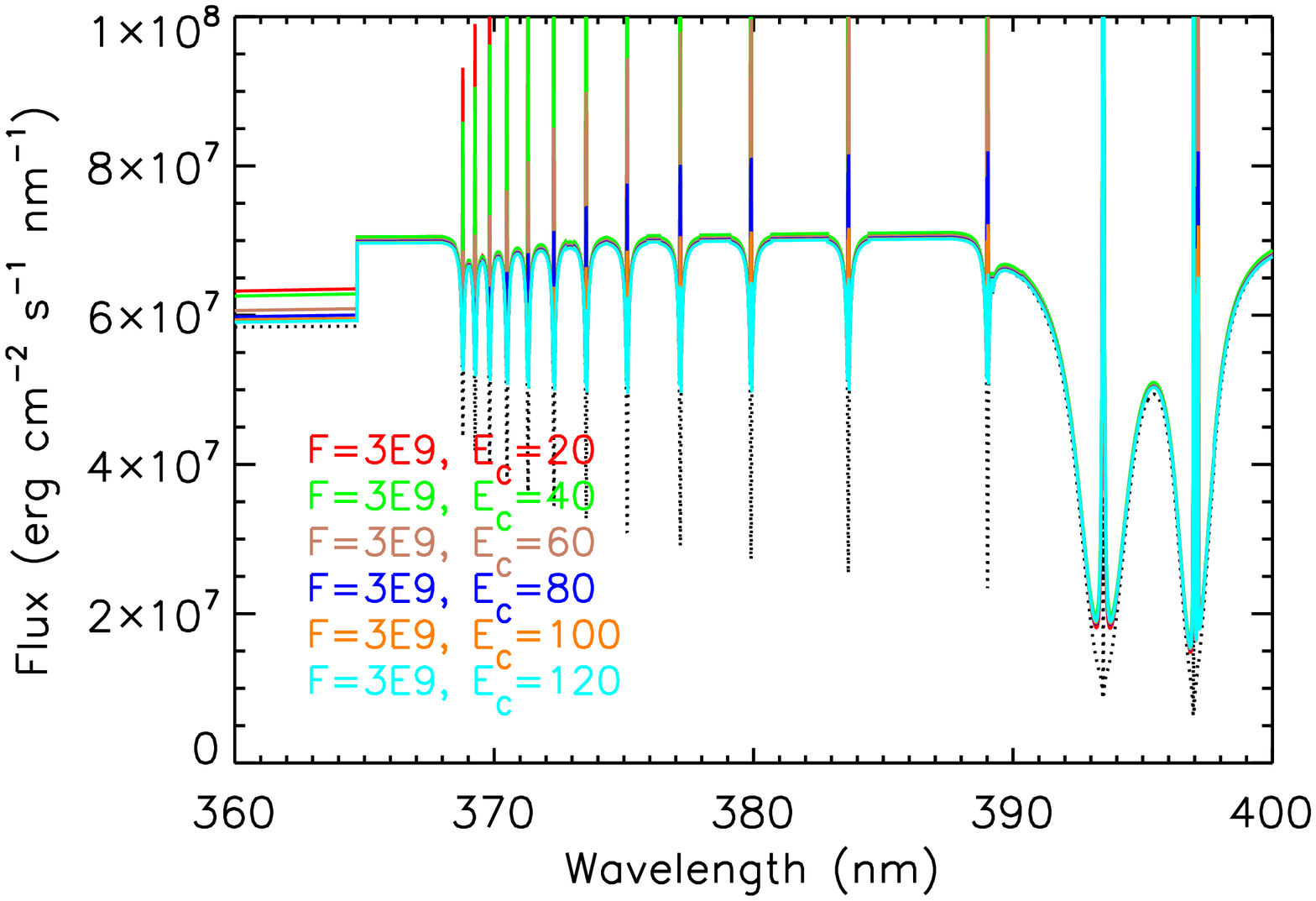}\\
    \includegraphics[width=0.49\linewidth]{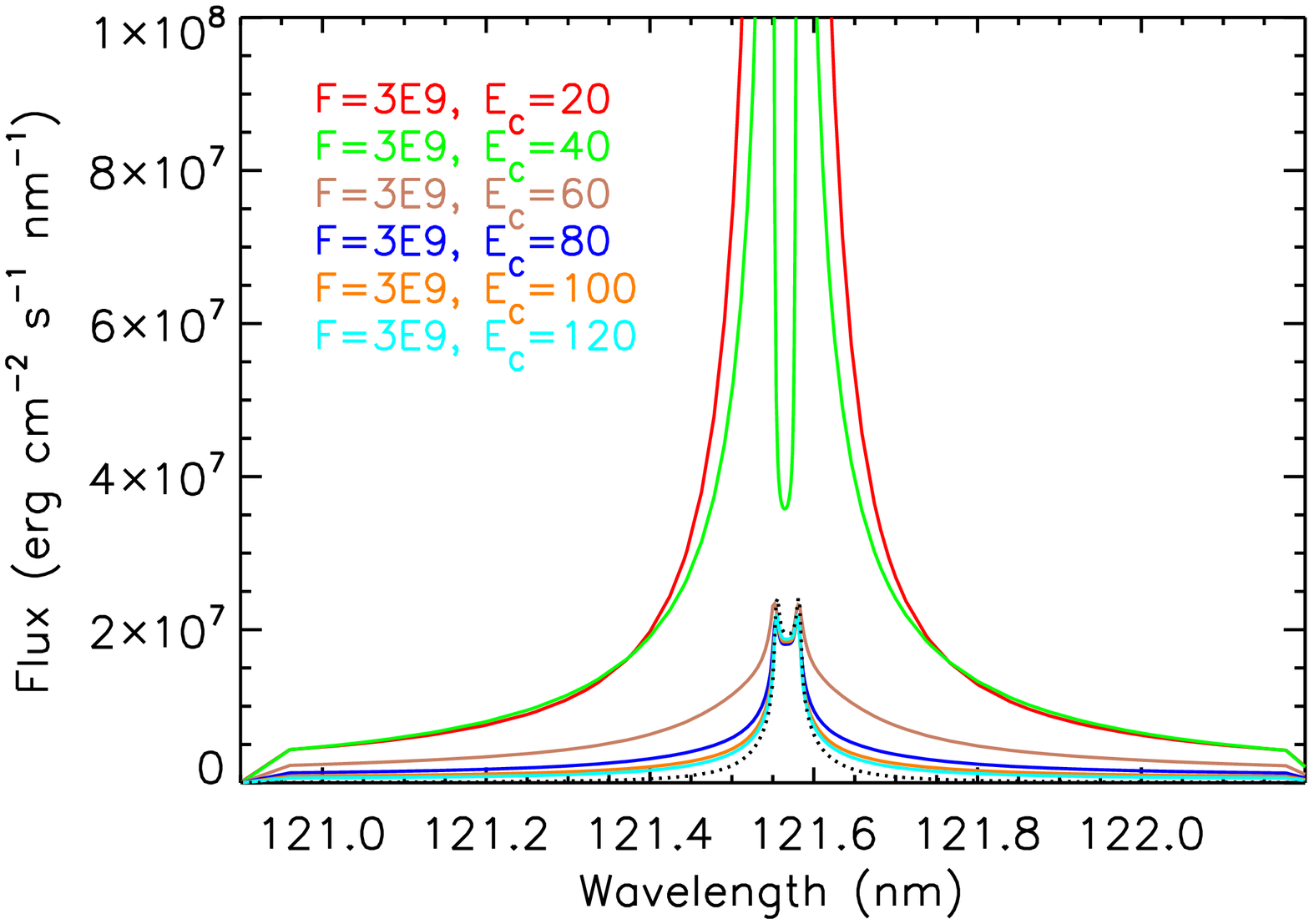}
    \caption{RH synthetic spectra from the electron beam-driven models with a beam flux of  $3\times10^9$ erg cm$^{-2}$ s$^{-1}$. The dotted line marks the quiescent profiles.}
    \label{fig:synthetic_sp}
\end{figure}

Figure \ref{fig:synthetic_sp} shows the RH synthetic spectra for the electron beam-driven models with a flux of $3\times10^{9}$ erg cm$^{-2}$ s$^{-1}$ in the wavelength range that covers the higher order Lyman lines and LyC, the Ly$\alpha$, the higher order Balmer lines and BaC. The figure shows a qualitative difference between the Lyman and Balmer emission. The Balmer emission, which is especially well illustrated on the BaC level, gradually decreases with increasing E$_C$, the LyC is only detected for low values of E$_C$. Figure \ref{fig:4} shows that for beam fluxes less than $1\times10^{10}$ erg cm$^{-2}$ s$^{-1}$ we can find E$_C$ within the modelled range, for which the excess radiative losses in the LyC are almost zero. In the spectrum (Figure \ref{fig:synthetic_sp}) this is manifested as no excess emission in the LyC and no rise of the Ly$\alpha$ peak emission.

\begin{figure}
    \centering
    \includegraphics[angle=90,width=0.6\linewidth]{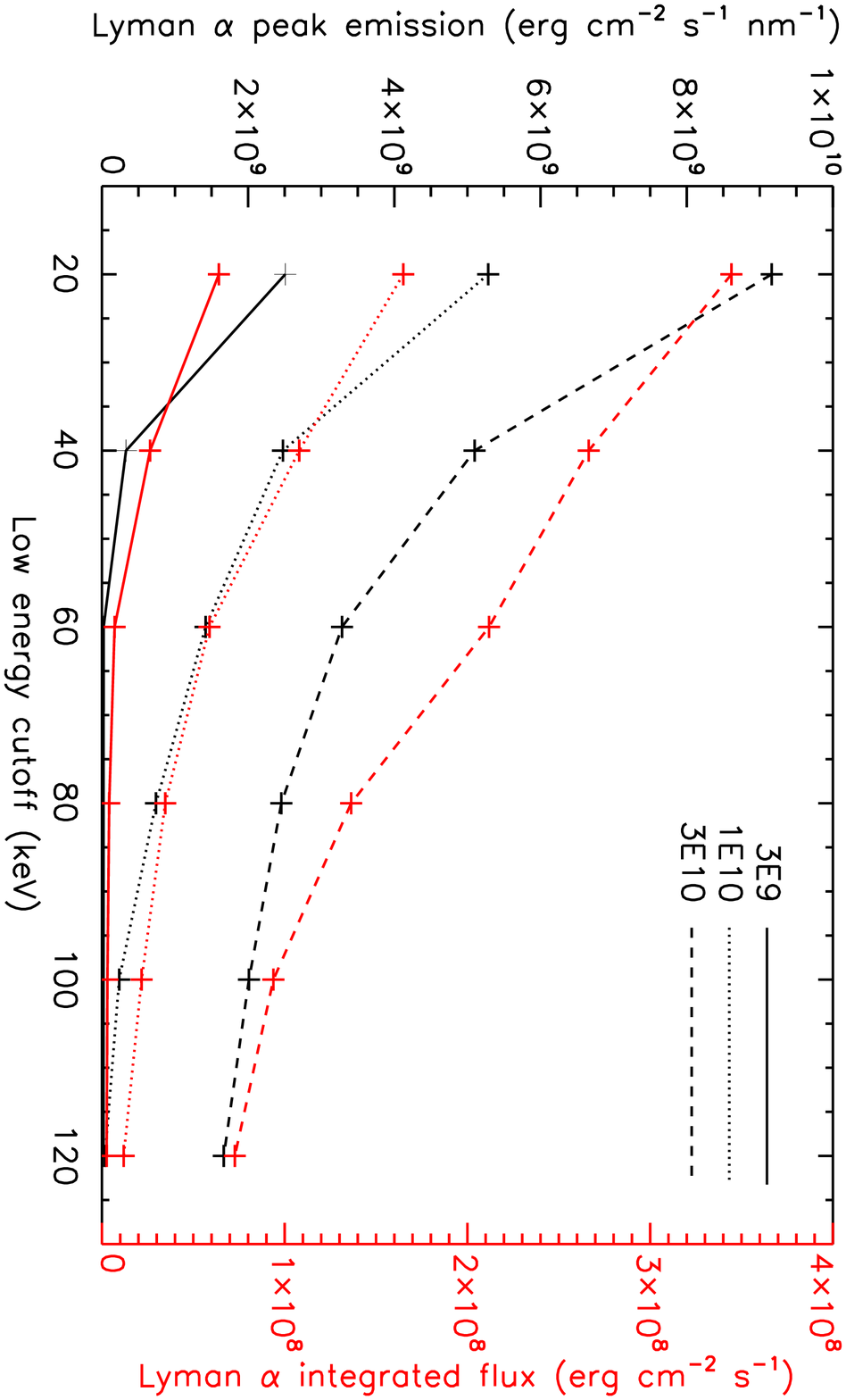}
    \caption{The spectral characteristics of Lyman $\alpha$ line for a range of E$_C$ and beam flux equal to $3\times10^{10}$ erg cm$^{-2}$ s$^{-1}$. The line flux was integrated over 1.2 nm with the line core being in the centre of the integrated wavelength region.}
    \label{fig:lya}
\end{figure}

\section{Discussion and concluding remarks}
We have used radiative hydrodynamic models to study the atmospheric response to electron beam-driven heating. Our analysis has shown that for sufficiently low electron beam fluxes and sufficiently high E$_C$ the flaring atmosphere does not produce any significant excess emission in the Lyman lines and continuum. These beam parameters still produce excess emission in the higher order hydrogen transitions, such as the Balmer or Paschen continua. This is due to the different atmospheric layers where the emission originates. The LyC is produced at the top of the chromosphere, which during flares is disturbed by the low energy particles, whereas the higher order continua are produced deeper in the atmosphere, and their excess emissions therefore appear in all our models (Figure \ref{fig:1}). The particle beams with the high E$_C$ do not contain the low energy particles that would be stopped at the top of the chromosphere, hence they do  not yield excess flare emission in the LyC. The lack of a response in the LyC can be considered as an intriguing result as the type II flare that our models are based on has a GOES X1.0 classification and hence a strong X-ray emission.

Flares with no Lyman emission have not been reported in the literature. This may be due to the lack of suitable observational data and the sparsity of flare beams with sufficiently high E$_C$ (\citealt{Fletcher:2007aa}). \cite{Milligan:2014aa} published an extensive multi-instrument analysis of a type I X-class solar flare and found that the Ly$\alpha$ dominated the measured radiative losses. Their upper estimates of the E$_C$ and the beam flux were in a range of 21.8 - 25.9 keV and $2.7\times10^{10} - 4.7\times10^{11}$ erg cm$^{-2}$ s$^{-1}$, respectively, therefore more energetic than those studied in our work. We speculate that they found the Ly$\alpha$ to dominate the radiative losses, because its core is formed at the top chromosphere and transition region (\citealt{Vernazza:1981aa}), and therefore is not  subjected to significant re-absorption by the  overlying layers of the atmosphere. On the other hand the continua are formed in the deeper layers of the atmosphere, and even if a significant portion of energy is emitted in these transitions, only a fraction of the energy can be detected due to re-absorption. Indeed, they estimate that the detected energy in the EUV H I, He I, He II continua, the He II (30.4 nm) and Ly$\alpha$ lines, the UV continua at 160 nm and 170 nm, the WL continua at 450.4 nm, 555.0 nm, and 668.4 nm, and the Ca II H line (396.8 nm) accounted for not more than 15\% of the total energy delivered to the atmosphere by the electron beams. The Spectral Investigation of the Coronal Environment (SPICE, \citealt{2013SPIE.8862E..0FF}) instrument on board the Solar Orbiter, that is due to be launched in 2020, will record spectra in the wavelength ranges $70.4 - 79.0$ nm and $97.3 - 104.9$ nm. These spectral ranges cover the higher Lyman lines and continuum as well as lines from several ionized species formed at temperatures from 10 thousand to 10 million K.\\
Particle beams with a E$_C$ greater than 100 keV were directly observed only once by \cite{Warmuth:2009aa}, but they were also found to be consistent with the spectral features of type II WLF \citep{Prochazka:2018aa}. Their work confirms that these beams produce a weaker excess Balmer emission than the low E$_C$ particle beams consistent with the definition of the type II WLFs.
 From an observational point of view it is easier to detect the flare excess in the Lyman lines and continuum due to the absence of a strong background spectrum. 
 Our work shows that the absence of an excess emission in the LyC and no increase of the Ly$\alpha$ peak emission are the best indicators of the type II WLF and for the first time reveals their qualitative difference with the type I events. The definition of type I and type II white-light flares has been traditionally based on the observational signatures that emanate from the hydrogen Balmer lines and continuum (\citealt{Machado:1986aa}). The IRIS mission has acquired a large number of flare datasets that provide a wealth of upper chromospheric and transition region diagnostics that may also be used to discriminate between these two type of flare events. 
 
\acknowledgments
The research leading to these results has received funding from the European Community's Seventh Framework Programme (FP7/2007-2013) under grant agreement no. 606862 (F-CHROMA). We would like to thank Joel Allred, Mats Carlsson, Adam Kowalski and Han Uitenbroek for the development of the numerical codes used in this study.


\end{document}